\documentclass{article}
\newcommand{\bc}{\begin{center}}
\newcommand{\ec}{\end{center}}
\newcommand{\Pb}{\hbox{{I}\kern-.1667em\hbox{P}}}
\newcommand{\Ex}{\hbox{{I}\kern-.1667em\hbox{E}}}
\newcommand{\Real}{\hbox{{I}\kern-.1667em\hbox{R}}}
\newcommand{\sReal}{\hbox{\scriptsize {I}\kern-.1667em\hbox{R}}}
\newcommand{\bfo} {\mbox{\boldmath $\omega$}}

\newcommand{\bfvr} {\mbox{\boldmath $\varrho$}}

\usepackage{amsthm,amsmath,amssymb,graphicx,subfig}
\title{\textbf{Regularizing Portfolio Risk Analysis:\\ A Bayesian Approach}}
\author{Sourish Das$^a$, Aritra Halder$^a$ and Dipak K. Dey$^b$\\
$^a$ Department of Mathematics, Chennai Mathematical Institute, India\\
$^b$ Department of Statistics, University of Connecticut, USA}
\date{}

\begin{document}
\maketitle

\vspace{.1 in}

\newtheorem{theorem}{Theorem}
\newtheorem{result}{Result}

\vspace{.2 in}

\begin{abstract}
It is important for a portfolio manager to estimate and analyse recent portfolio volatility to keep the portfolio's risk within limit. Though the number of financial instruments in the portfolio can be very large, sometimes more than thousands, daily returns considered for analysis are only for a month or even less. In this case rank of portfolio covariance matrix is less than full, hence solution is not unique. It is typically known as the ``ill-posed" problem. In this paper we discuss a Bayesian approach to regularize the problem. One of the additional advantages of this approach is to analyze the source of risk by estimating the probability of positive `conditional contribution to total risk' (CCTR). Each source's CCTR would sum up to the portfolio's total volatility risk. Existing methods only estimate CCTR of a source, and does not estimate the probability of CCTR to be significantly greater (or less) than zero. This paper presents Bayesian methodology to do so. We use a parallelizable and easy to use Monte Carlo (MC) approach to achieve our objective. Estimation of various risk measures, such as Value at Risk and Expected Shortfall, becomes a by-product of this Monte-Carlo approach.
\end{abstract}
\newpage
\section{Introduction}\label{Intro}
The recent euro-zone crisis remind us that `risk analysis' is always an essential part of the theory of portfolio management. Markowitz(1952) \cite{Markowitz1952}, first proposed volatility (or standard deviation) as the risk measure. As volatility provides an idea about average loss (or gain) of portfolio; more new risk measures like `\textit{Value at Risk}' (VaR) and  `\textit{Expected Short Fall}' (ESF)  for extreme losses are also popular. Although Basel II regulatory framework requires inclusion of VaR and ESF, till date volatility plays an important role in finance. For example, volatility can be traded directly on the open market through \textit{Exchange Traded Fund} of VIX index and indirectly through derivatives.

It is essential to measure the volatility and identify the main sources of volatility. Often the number of financial instruments of well diversified mutual funds or pension funds are more than thousands. Such funds invest in foreign countries on a regular basis; sometimes in more than fifty to sixty different countries. However, portfolio managers are concerned about the stationarity in long time-series data and interested about only recent volatility which consider daily returns of a month and sometimes even less. As the number of sectors, or countries, or components, of a portfolio is greater than the number of days of return, the rank of the portfolio covariance matrix is less than full; such cases yield non-unique solutions. Generally it is known as the ``ill-posed" problem.

In this paper, we mainly focus on estimating and analysing the sources of portfolio volatility under ``ill-posed" condition. We address the mean-variance optimization problem which admits an analytical solution for the optimal weights of a given portfolio. This optimization procedure requires an estimate of the portfolio covariance matrix. But due to the ``ill-posed" structure of the problem, regular sample covariance would not work here. Therefore, we propose a regularized plug-in Bayes estimator for the portfolio covariance matrix and use the optimized weights for the given portfolio. Using this setup we evaluate its out-sample performance using a Monte Carlo Algorithm.

In a 3-series paper, Ledoit and Wolf (2003, 2004a, 2004b)\cite{LedoitWolf2003,LedoitWolf2004a,LedoitWolf2004b} showed the use of Shrinkage estimator on actual stock market data keeping all other steps of optimization process the same. By doing so they reduced the tracking error relative to a ad-hoc index. As a result it substantially increased the realized information ratio of active portfolio managers. Ledoit and Wolf (2004a) suggested a distribution-free approach to regularizing the covariance matrix, in this paper however we impose a probability structure on the covariance matrix for obvious reasons. The proposed regularized plug-in Bayes estimator in this paper bears a direct relationship with the empirical Bayes estimator suggested by  Ledoit and Wolf (2004a) \cite{LedoitWolf2004a}, for the covariance matrix. 
\newline
Also, Golsnoy and Okhrin (2007)\cite{GolosnoyOkhrin2007} showed the improvement of portfolio selection by using the multivariate shrinkage estimator for the optimal portfolio weights. Recently, Das and Dey (2010) \cite{DasDey2010} introduced some Bayesian properties of multivariate gamma distribution for covariance matrix. In this paper we use this Bayesian approach to regularize the estimation problem. Under certain conditions, the posterior distribution of the portfolio covariance matrix is proper and has a closed form inverted multivariate gamma distribution. Consequently, the solution of covariance matrix is unique. The rest of the article is organized as follows. In Section \ref{PostCov}, we discuss the posterior distribution of portfolio covariance matrix and the condition under which it is proper. In Section \ref{RiskAna}, we present a parallelizable Monte Carlo algorithm to obtain posterior inference on the risk. In Section \ref{EmpStudy}, we demonstrate the method with two empirical data sets. The methodology of inference is applied  initially to a small dataset, consisting of different asset classes. Next we consider a portfolio consisting of the stocks from ``National Stock Exchange of India" (NSEI). Section \ref{Concl} concludes the paper.

\section{Posterior Distribution of the Portfolio Covariance Matrix\label{PostCov}}

Suppose, $S$ is a real symmetric sample portfolio covariance matrix of order $p$ with $\frac{p(p+1)}{2}$ variables $s_{ij}$. $\Sigma=((\sigma_{ij}))$ is the corresponding population portfolio covariance matrix, such that for a diagonal matrix $D$ with diagonal elements $1$ and $-1$, $(D \Sigma D)^{-1}$ has non-positive off-diagonal elements. Hence due to Bapat's condition, Bapat (1989) \cite{Bapat1989}, $S$ with characteristic function as
$$
\psi_S(T)=E[i\exp\big\{\mathrm{tr}(TS)\big\}]=|I_p-i\beta \Sigma T|^{-\alpha},
$$
has the density function  
$$
f(S)=\frac{|\Sigma|^{\alpha}}{\Gamma_p(\alpha)\beta^{\alpha p}}\exp\Big\{-\frac{1}{\beta}\mathrm{tr} \left ( \Sigma^{-1}S \right ) \Big\}|S|^{\alpha-\frac{1}{2}(p+1)},~~ S>0
$$
has an infinitely divisible multivariate gamma distribution with parameters $\alpha \geq \frac{p-1}{2}$, $\beta \geq 0$ and $\Sigma$ a positive definite matrix. Note that if $0 \leq \alpha \leq \frac{p-1}{2}$, $S$ has a degenerate distribution. If we choose $\alpha=\frac{n-1}{2}$ and $\beta=2$ then $S$ follows a Wishart distribution, i.e., $S \sim \mathcal{W}(n-1,\Sigma)$ (see Anderson, (1984) \cite{Anderson1984}, pp 252). If $p \geq n$, then $S$ is less than full rank, and the sampling distribution of $S$ is degenerate and that is no valid statistical inference can be implemented for such cases.
\newline
Das and Dey (2010) \cite{DasDey2010} showed that if $\Sigma$ has prior as inverted multivariate gamma distribution, i.e., if $\Sigma \sim \mathcal{MG}^{-1}_p(a,\beta,\Psi)$ then the posterior distribution of $\Sigma$ is 
$$
\Sigma|S \sim \mathcal{MG}^{-1}_p(\alpha+a,\beta,S+\Psi).
$$
Note that as long as $(\alpha + a) \geq \frac{p-1}{2}$, the posterior distribution is proper. Suppose $n \leq p$, i.e., $\alpha \leq \frac{p-1}{2}$, where $\alpha=\frac{n-1}{2}$; then the sampling distribution of $S$ is degenerate. However, if we choose the prior degrees of freedom parameter $a$ to be such that 
$$
\alpha + a \geq \frac{(p-1)}{2}
$$
that is $a \geq \frac{p-n}{2}$ then the posterior distribution of $\Sigma$ is proper. Hence, we will be able to carry out Bayesian Inference.
\newline
If we choose $a=\frac{n_0}{2}$ and $\beta=2$ then the prior of $\Sigma$ is inverse Wishart distribution,   
\begin{equation}
\Sigma \sim \mathcal{W}^{-1}(n_0,\Psi) \label{WishartPrior}
\end{equation}
and posterior distribution of $\Sigma$ is 
\begin{equation}
\Sigma|S \sim \mathcal{W}^{-1}(n_0+n-1,\Psi+S) \label{WishartPosterior}
\end{equation}
for details see Anderson (1984) \cite{Anderson1984}. 

\bigskip

The 3-series paper of Ledoit and Wolf (2003, 2004a, 2004b), established the reasons for the sample covariance matrix failing to provide a good estimate for the portfolio covariance structure, and showed the need for regularization even when the problem is not ``ill-posed". However, there was a lack of scope for implementing inferential procedures on their structural framework, regarding quantities like \textit{Conditional Contribution to Total Risk}. The main reason behind this, being the problem of assigning a suitable model for the anticipated return distribution. In this paper an assumption for the covariance matrix $S \sim \mathcal{W}$, has an underlying assumption of normal/normal-component-mixture distribution on the anticipated returns. The justification for the assumption $S \sim \mathcal{W}$ is provided by the Bayesian formulation as in Gelman et al. \cite{Gelman2004}. The marginal posterior distribution for anticipated return would be $t$. This would provide a superior apprach to modelling the anticipated return in comparision to using a normal distribution.

\bigskip

If $n<p$, then we choose the prior degrees of freedom as 
\begin{equation}
n_0=(p-n)+c.\label{EqChooseDf}
\end{equation}
for $c>0$. This ensures posterior distribution to be proper. The posterior mode of $\Sigma$ is
\begin{eqnarray}
M(\Sigma|S)&=&\frac{\Psi+S}{n_0+n+p} \nonumber\\ 
&=&\frac{n_0+p+1}{n_0+n+p}.\frac{\Psi}{n_0+p+1}+\frac{n-1}{n_0+n+p}.\frac{S}{n-1}\nonumber\\
&=& q\frac{\Psi}{n_0+p+1}+(1-q)\frac{S}{n-1}, \label{SMode}
\end{eqnarray}
where $q=~\frac{n_0+p+1}{n_0+n+p}$.
Clearly posterior mode of $\Sigma$ is a shrinkage estimator which is a weighted average of prior distribution's mode and sample covariance estimator. Das and Dey (2010) \cite{DasDey2010} showed posterior mode is also a Bayes estimator under a Kullback-Leibler type loss function. Therefore under properly chosen prior degrees of freedom ($\alpha$ or $n_0$) and positive definite $\Psi$, posterior mode of $\Sigma$ is also a Bayes shrinkage estimator which regularizes the solution. The posterior distribution of $\Sigma$ is proper, which helps us to regularize the portfolio optimization procedure while conducting our empirical study in Section \ref{EmpStudy}.

\bigskip

Under the assumption of $\mathrm{E}[S]=\Sigma$, where $S$ is the usual sample portfolio covariance matrix and $\Psi=\mathrm{I}$; Ledoit and Wolf (2004a) \cite{LedoitWolf2004a} showed that the folowing estimator,
\begin{equation}
\Sigma_{1}\; =\; \frac{\beta^2}{\delta^2}\mu \mathrm{I}+\frac{\alpha^2}{\delta^2}S \label{LedoitEst}
\end{equation}
is asymptotically optimal under the Frobenius norm. In this work we assume $S \sim \mathcal{W}(n-1,\Sigma)$, which implies $\mathrm{E}\Big(\frac{S}{n-1}\Big)=\; \Sigma$. Hence, the corresponding estimator is 
\begin{equation}
\Sigma_{2}~ =~ \frac{\beta^2}{\delta^2}\mu \mathrm{I}+\frac{\alpha^2}{\delta^2}\frac{S}{n-1}. \label{DHDEst1}
\end{equation}
Now we consider the definition of functional order from Knuth(1976) \cite{Knuth1976}.
\newline
\textbf{Definition:} For functions $f$ and $g$, let $O$, $\Omega$ and $\Theta$ be defined as
\begin{itemize}
\item $O(f(x))~=~\lbrace g(x) \Big| \exists ~ c>0, ~ x_0 \in \mathbb{R}^{+} ~  |g(x)|\leq cf(x) ~ \forall x \geq x_0\rbrace$,
$O(f(x))$ is the set of all such functions whose magnitude is upper-bounded by constant times $f(x)$.
\item $\Omega(f(x))=\lbrace g(x) \Big| \exists ~ c>0, ~ x_0 \in \mathbb{R}^{+} ~ |g(x)|\geq cf(x) ~\forall x \geq x_0\rbrace$,
$\Omega(f(x))$ is the set of all such functions whose magnitude is lower-bounded by constant times $f(x)$.
\item $\Theta(f(x))=\lbrace g(x) \Big| \exists ~ c_1, c_2>0, ~ x_0 \in \mathbb{R}^{+}~ c_1f(x) \leq |g(x)|\leq c_2f(x) ~ \forall x \geq x_0\rbrace$,
$\Theta(f(x))$ is the set of all such functions whose magnitude is lower as well as upper-bounded by constant/s time/s $f(x)$.
\end{itemize}

\textbf{Note 1:} Now we will check if the co-efficient of the first term in (\ref{SMode}) and $\frac{\beta^2\mu}{\delta^2}$ in (\ref{DHDEst1}) are of the same functional order. This facilitates a measure for the amount of shrinkage towards the target. In (\ref{SMode}), the shrinkage target is the prior, while in (\ref{LedoitEst}) and (\ref{DHDEst1}), it is $\mu\mathrm{I}$
\newline
In the equation (\ref{DHDEst1}) $\mu= \left < \Sigma, \mathrm{I}\right > \;=\; \frac{\mathrm{tr} \left (\Sigma \mathrm{I}\right )}{p}$. Here, $\alpha^2=||\Sigma-\mu\mathrm{I}||^2$, $\beta^2=\mathrm{E}\Big[\Big|\Big|\frac{S}{n-1}-\Sigma\Big|\Big|^2\Big]$ and $\delta^2=\mathrm{E}\Big[\Big|\Big|\frac{S}{n-1}-\mu\mathrm{I}\Big|\Big|^2\Big]$. Due to the imposed  probability structure on the covariance matrix $S$, we have the following results.
\begin{result}
$\alpha^2= ||\Sigma-\mu\mathrm{I}||^2= \frac{1}{p}\Big[\mathrm{tr}(\Sigma^2)-\frac{[\mathrm{tr}(\Sigma)]^2}{p}\Big]=\; \frac{1}{p^2}\Big[(p-1)\sum_{i=1}^{p}\sigma_{ii}^2+\mathop{\sum \sum}_{i\neq j}^{p} \sigma_{ii}\sigma_{jj}(p\;\rho_{ij}^2-1)\Big]$
\end{result}

\begin{result}
$\beta^2=\mathrm{E}\Big[\Big|\Big|\frac{S}{n-1}-\Sigma\Big|\Big|^2\Big] = \frac{2}{n-1}\Big[\frac{1}{p}\sum_{i=1}^{p}\sigma_{ii}^2\Big]$
\end{result}

\begin{result}
$\delta^2 = \alpha^2+\beta^2=\; \frac{1}{p}\Big\lbrace\mathrm{tr}(\Sigma^2) -\frac{[\mathrm{tr}(\Sigma)]^2}{p}\Big\rbrace+\frac{2}{n-1}\Big[ \frac{1}{p}\sum_{i=1}^{p}\sigma_{ii}^2\Big]$
\end{result}

\begin{result}
$\mu = \frac{1}{p}\sum_{i=1}^{p}\sigma_{ii}$.
\end{result}

Result 4 follows from the definition of Fr$\ddot{o}$benius norm and the proofs of Result 1,2 and 3 are presented in Section \ref{App}. 
\newline
Now we examine the denominator of $\frac{\beta^2\mu}{\delta^2}$ in (\ref{DHDEst1}),
\begin{eqnarray}
\delta^2 &=& \frac{1}{p}\Big\lbrace\mathrm{tr}(\Sigma^2) -\frac{[\mathrm{tr}(\Sigma)]^2}{p}\Big\rbrace+\frac{2}{n-1}\Big[ \frac{1}{p}\sum_{i=1}^{p}\sigma_{ii}^2\Big], \nonumber \\
\delta^2 &=& \frac{1}{p^2}\Big[(p-1)\sum_{i=1}^{p}\sigma_{ii}^2+\mathop{\sum \sum}_{i\neq j}^{p}\sigma_{ii}\sigma_{jj}(p\;\rho_{ij}^2-1)\Big]+\frac{2}{n-1}\Big[\frac{1}{p}\sum_{i=1}^{p}\sigma_{ii}^2\Big]. \label{DeltaEqn}
\end{eqnarray}
Multiplying both sides of equation \ref{DeltaEqn} by $(n-1)p^2$ we get,
\begin{eqnarray}
(n-1)p^2 \; \delta^2 &=& \Big\lbrace 2p+(n-1)(p-2)+(n-1)\Big\rbrace \sum_{i=1}^{p}\sigma_{ii}^2+ (n-1) \mathop{\sum \sum}_{i\neq j}^{p}\sigma_{ii}\sigma_{jj}\Big(p\rho_{ij}^2-1\Big)\label{Numer}\\
&=& \Big\lbrace 2p+(n-1)\Big\rbrace \sum_{i=1}^{p}\sigma_{ii}^2+ (n-1) \Big[\mathop{\sum \sum}_{i\neq j}^{p} \sigma_{ii}\sigma_{jj}\Big(p\rho_{ij}^2-1\Big)+(p-2)\sum_{i=1}^{p}\sigma_{ii}^2 \Big]\nonumber\\
&=& A_1+A_2, \nonumber
\end{eqnarray}
where, 
\begin{eqnarray*}
A_1 &=& \Big\lbrace 2p+(n-1)\Big\rbrace \sum_{i=1}^{p}\sigma_{ii}^2 
\end{eqnarray*}
is a function of the variances, and 
\begin{eqnarray*}
A_2 &=& (n-1) \Big[\mathop{\sum \sum}_{i\neq j}^{p}\sigma_{ii}\sigma_{jj}\Big(p\rho_{ij}^2-1\Big)+(p-2)\sum_{i=1}^{p}\sigma_{ii}^2 \Big]
\end{eqnarray*}
is a function of the variances and the covariances.
\newline
We consider $(p\rho_{ij}^2-1)$ from $A_2$, and if
\begin{eqnarray*}
(p\rho_{ij}^2-1) &\geq& 0 \hspace{.5cm} , p \geq 2\\
\Rightarrow \rho_{ij}^2 &\geq& \frac{1}{p} \hspace{.5cm} , p \geq 2\\
\end{eqnarray*}
which reduces to $\rho_{ij}^2 \geq 0$, as $p \to \infty$, which implies $A_2 \geq 0$. Now, we consider $(n-1)p^2\delta^2$ from (\ref{Numer})
\begin{eqnarray}
(n-1)p^2\delta^2 &=& A_1+A_2 \geq A_1 = (n+2p-1) \sum_{i=1}^{p}\sigma_{ii}^2 \nonumber\\
&\geq& (n+p)\Big\lbrace p \; \mathop{\mathrm{min}}_{i}(\sigma_{ii}^2 )\Big\rbrace\nonumber\\
&\Rightarrow& (n-1)p^2\delta^2 = \Omega\Big((n+p)p\Big).\label{DenomO}
\end{eqnarray}
Therefore, (\ref{DenomO}) provides a lower-bound for the denominator of $\frac{\beta^2\mu}{\delta^2}$. Now we consider the form of $\beta^2\mu$, from Results 2 and 4,
\begin{eqnarray}
(n-1)p^2\beta^2\mu&=& 2\Big[\sum_{i=1}^{p}\sigma_{ii}^2\Big]\Big[\sum_{i=1}^{p}\sigma_{ii}\Big]\label{VarSt}\\
&\leq& 2p^2 \Big\lbrace \mathop{\mathrm{max}}_{i} \sigma_{ii}^2\Big\rbrace \Big\lbrace \mathop{\mathrm{max}}_{i} \sigma_{ii}\Big\rbrace\nonumber\\
&\Rightarrow& (n-1)p^2\beta^2\mu = O(p^2).\nonumber\\
\end{eqnarray}
It provides an upper-bound for the numerator of $\frac{\beta^2\mu}{\delta^2}$. Hence we combine these findings in the following theorem.
\begin{theorem}
If S, is the sample portfolio covariance matrix, such that $S \sim \mathcal{W}(n-1,\Sigma)$, then for (\ref{SMode}) and (\ref{DHDEst1}) we have,
$$
O\Big(\frac{\beta^2\mu}{\delta^2}\Big)=O\Big(\frac{q\mu}{n_0+p+1}\Big)=O\Big(\frac{p}{n+p}\Big)
$$
provided $n_0=~O(p)$.
\end{theorem}
Theorem 1 establishes the functional equivalence of the distribution-free approach and the Bayesian approach. Note that from (\ref{SMode})
\begin{eqnarray}
\frac{q\mu}{n_0+p+1}=\; \Theta\Big(\frac{p}{n+p}\Big).\label{OrderBayes}
\end{eqnarray}
\textbf{Note 2: } Theorem 1 implies that the `degree of shrinkage' towards the target for (\ref{LedoitEst}), is less than a constant times $\frac{p}{n+p}$; whereas in (\ref{SMode}), the degree of shrinkage towards the target, is exactly $\frac{p}{n+p}$.

\bigskip

Equation (\ref{VarSt}) is a function of the variances, therefore we propose the following modification to the regularization,
\begin{eqnarray*}
\Sigma_{3}&=& \rho \lambda^{\prime} \mathrm{I} + (1-\rho) \frac{S}{n-1},
\end{eqnarray*}
where, $\lambda = (s_{11},\ldots,s_{pp})^{\prime}$. We have two choices of weights for the regularization. Using (\ref{OrderBayes}), we can choose $\rho$ as the Bayesian weights and compare the performance with the asymptotic weights for $\rho$ presented in  Ledoit and Wolf (2003) \cite{LedoitWolf2003}. The asymptotic weight $\rho$ is obtained by minimizing $\mathrm{E}[L(\rho)]$, under the squared error loss, that is 
\begin{eqnarray*}
L(\rho)&=& \Big|\Big| \hat{\Sigma_{3}}-\Sigma \Big|\Big|^2 \hspace{1cm} \text{subject to } \rho \geq 0.
\end{eqnarray*}
On solving the optimisation problem as in Section \ref{App}, under the probability structure imposed on $S$, we have
\begin{result}
$$
\rho \; =\;  \Big[\frac{[(n-2)^2+1]\sum_{i=1}^{p}\sigma_{ii}^2-(n-2)\sum_{i=1}^{p}\sigma_{ii}-(n-1)\sum_{i=1}^{p}\sum_{j=1}^{p}\sigma_{ij}^2}{(n-2)^2[2\sum_{i=1}^{p}\sigma_{ii}^2+\sum_{i=1}^{p}\sigma_{ii}]}\Big].
$$
\end{result}
We compare the performance of the resulting optimal estimators under the different weights, the first being
\begin{eqnarray}
\Sigma_{13}&=& \rho \lambda^{\prime} \mathrm{I} + (1-\rho) \frac{S}{n-1}\label{lw-weight},
\end{eqnarray}
with $\rho$ as in Result 5, and the Bayesian Shrinkage Estimator
\begin{eqnarray}
\Sigma_{23}&=& \frac{\lambda^{\prime} \mathrm{I} + S}{n_0+n+p} \nonumber\\
&=& \frac{q}{n_0+p+1}\lambda^{\prime}\mathrm{I} + \frac{(1-q)}{n-1}S.\label{bayes-wt}
\end{eqnarray}

\section{Bayesian Inference for Analysing Risk\label{RiskAna}}

Ledoit and Wolf (2004a) \cite{LedoitWolf2004a} presented asymptotic properties of the regularized covariance estimator, under a distribution free approach. The lack of any distributional assumptions, prevents us from applying any inferential procedures for estimating quantities of interest, for example `Marginal Contribution to Total Risk' (MCTR) and `CCTR'. The `MCTR' and `CCTR' of individual securities provide justification for the risk-behaviour of the portfolio.   However, the distribution-free approach bypassed the need to model stock returns explicitly, and provided a justification for the method of regularization. In an attempt to approach a better regularization procedure, Ledoit and Wolf (2003) \cite{LedoitWolf2003} exploits the dependence structure of individual securities traded in a market, with the market index. This approach resulted in a market specific regularization, by altering the shrinkage target. In Bayesian terminology this indicates that the prior information in comparision to Ledoit and Wolf (2004a) is altered.

From, the applicability point of view, the downside of this approach, was the computationally intensive calculation of the weights, for the regularization procedure of the covariance matrix. As we have seen in Section \ref{PostCov}, the theorem, provides a simple set of weights, showing functional equivalence with the posterior mode, of a shrinkage estimator under the minimal assumptions of a Wishart probability structure on the covariance matrix. Moreover the new regularization procedure suggested for the covariance matrix, does not utilise any additional correlation structure with the market index. This provides a more generalized and implementable approach to regularize the covariance matrix.

Under the assumption of no `short-selling', the performance of Markowitz's \cite{Markowitz1952} expectation-variance optimisation procedure with $n << p$, has been known to falter due to singularity conditions already stated in the Section \ref{Intro}. Regularizing, the covariance matrix provides a bypass to the problem. $\Sigma$, being ill-posed implies $\forall \; \bfo \in \mathbb{R}^{p},~ \bfo^{T}\Big\lbrace \frac{S}{n-1}\Big\rbrace \bfo \geq 0$. For a strictly positive definite prior-information matrix $\lambda^{\prime} \mathrm{I_{p}}$
\begin{eqnarray*}
\bfo^{T}M(\Sigma|S)\bfo &=&  \bfo^{T}\Big(\frac{q}{n_0+p+1}\lambda^{\prime} \mathrm{I_{p}}~+~(1-q)~\frac{S}{n-1}\Big)\bfo \hspace{1cm} (q > 0)\\
&=& \bfo^{T}\Big(\frac{q}{n_0+p+1}\lambda^{\prime} \mathrm{I_{p}}\Big)\bfo+\bfo^{T}\Big( (1-q)~\frac{S}{n-1}\Big)\bfo \; >\;  0, 
\end{eqnarray*}
$\forall \; \bfo \in \mathbb{R}^{p}$, since $\bfo^{T}\Big(\frac{q}{n_0+p+1}\lambda^{\prime} \mathrm{I_{p}}\Big)\bfo~>~0$.
The problem of maximizing the expectation and minimizing the variance as in Markowitz (1952), is transformed into the problem of utility maximization of the investor as in Fabozzi et.al(2008) \cite{Fabozzi2008}. Under the general setup for optimization, we have the expected utility function E as a function of the returns and the variance.
\begin{eqnarray*}
E&=&\mu-\frac{A}{2}\lambda,
\end{eqnarray*}
where $A$ is the increase in risk, the investor is willing to tolerate per unit increase in return. It serves as a relative risk aversion measure. We assume that the expected utility function is $0$, while proceeding with the optimization. Therefore, $A$ is a matrix of the relative risk aversion towards each security
\begin{eqnarray*}
2\frac{\mu_i}{\sigma_{i}^2} &=& \mathop{\mathrm{Diag}}_{i}(A) \hspace{1cm} \forall i= 1,\ldots,p
\end{eqnarray*}
which is a function of the Sharpe Ratio, for indvidual securities. The regularized covariance matrix is used to implement a quadratic-optimizer sub-routine to determine the weights for the optimal portfolio. Consequently, we can calculate the portfolio risk for the selected number of stocks in the portfolio.

Once portfolio risk has been calcuated using the weights, the next step is to evaluate the sources of risk and how they interrelate. There could be many different sources of risk, like individual stocks, sectors, asset classes, industries, currencies or style factors. Therefore, we keep the notion for \textit{sources of risk} generic.
\newline
We consider an investment period where $r_j$ denote the return of source $j$ for the same period, where $j=1,2,\ldots,p$. The anticipated portfolio return over the period is
$$
R_p=\sum_{j=1}^p \omega_j r_j,
$$
where $\omega_j$ is the portfolio's exposure to the source $j$, i.e., the portfolio weight, such that $\omega_j \geq 0$ and $\sum_{j=1}^p\omega_j=1$, see Ruppert (2004) \cite{DavidRuppert2004}. Portfolio volatility is defined as
$$
\sigma_p=\sqrt{\bfo^T \Sigma~ \bfo},
$$
where $\bfo^T=\{\omega_1,\omega_2,...,\omega_p\}$. The  portfolio manager determines the size of $\omega_j$ at the beginning of the investment period, typically using Markowitz-type optimization. Clearly, the  weights ($\omega_j$) play an important role as regulators of the portfolio's total  volatility, along with the covariance structure of the portfolio. We have already dealt with the issue of regularizing the portfolio's covariance structure. However, it is also important for a manager to quantify, how sensitive the portfolio volatility is with respect to a small change in $\bfo$. This can be achieved by differentiating the volatility with respect to weight, i.e.,
$$
\frac{\partial(\sigma_p)}{\partial \bfo}=\frac{1}{\sigma_p}.\Sigma.\bfo=\bfvr,
$$
where $\bfvr=\{\varrho_1,\varrho_2,...,\varrho_p\}$, is known as the `Marginal Contribution to Total Risk' (MCTR), see Menchero \textit{etal.} (2011) \cite{MencherDavis2011} and Baigent (2014) \cite{Baigent2014}. Note that the MCTR for source $i$ is given by the following expression
$$
\varrho_i=\frac{1}{\sigma_p}\sum_{j=1}^p \sigma_{ij}~\omega_j.
$$
The CCTR is the amount that a source contributes to the total portfolio volatility. In other words, if $\zeta_j=\omega_j.\varrho_j$ is the CCTR of source $j$ then
\begin{eqnarray*}
\sigma_p&=&\sum_{j=1}^p\zeta_j=\sum_{j=1}^p\omega_j.\varrho_j.
\end{eqnarray*}
Therefore the total volatility can be viewed as weighted average of the MCTR. 

Now in order to estimate the MCTR and CCTR, a regularized estimate of $\Sigma$ is required, because the portfolio weights $\boldsymbol \omega$ are pre-determined by the manager. However, for all practical purposes a manager is more interested in estimating the $P(\varrho_j<0)$ or $P(\zeta_j<0)$. The reason being, $\varrho_j < 0$ or $\zeta_j < 0$ implies that the source $j$ reduces the total risk. In section \ref{PostCov}, we presented that the posterior distribution of $\Sigma$ follows $\mathcal{W}^{-1}(n_0+n-1,S+\Psi)$. Also, in order to estimate the $P(\varrho_j>0)$ we present a Monte Carlo (MC) method based on independent replications as follows:
\begin{itemize}
\item Step 1: For iteration $i$, generate a sample $\Sigma_{(i)}$ from $\mathcal{W}^{-1}(n_0+n-1,S+\Psi)$
\item Step 2: Compute $\sigma_{p}^{(i)}=\sqrt{\bfo^T\Sigma^{(i)}\bfo}$
\item Step 3: Compute $\bfvr^{(i)}=\frac{1}{\sigma_p^{(i)}}.\Sigma^{(i)}.\bfo$
\item Step 4: Compute $\zeta_j^{(i)}=\varrho_j^{(i)}\omega_j$ for all $j=1,2,...,p$.
\item Step 5: Set $i=i+1$ and go to Step 1.
\end{itemize}
Note that these are independent replications, that is all the steps of iteration $i$, does not depend on previous step $(i-1)$. If two parallel processors are available then iteration $i$ and $(i-1)$ can be implemented in parallel at the same time. In fact one can consider the algorithm to be an `embarrassingly parallel' algorithm (see Matloff (2011) \cite{Matloff2011}). Implementing the algorithm in parallel might not be required if $p$ is small, for example, if we consider four or five different asset classes as the risk sources. But if $p$ is very large, like for any mutual fund portfolio, number of stocks might be more than thousands. For a large p, generating $\Sigma^{(i)}$ will be slow and thus, consequent calculations in all the other steps will be slow as well. In such cases, parallelization of the algorithm is required to improve the time complexity for the algorithm.

Once the samples are generated, required MC statistics can be estimated easily, like 
\begin{equation}
P(\zeta_j>0)=\frac{1}{N}\sum_{i=1}^NI(\zeta_j^{(i)}>0),\label{EqProbCCTR}
\end{equation}
where $N$ is the simulation size, $I(A)=1$ if $A$ is true and $I(A)=0$ if $A$ is false.

\section{Empirical Study\label{EmpStudy}}

In this section we illustrate the methodology with two different sets of empirical data. Each exposes the inference procedure to a different practical situation. The first dataset, displays the performance of the algoritm on a small dataset, consisting of 5 asset classes and 3 months of monthly return for each asset class. The second dataset displays the performance of the algorithm on a large dataset, consisting daily log-returns of $p=450$ to $p=990$ stocks, over a period of 10 years.

\subsection{`Indices' Dataset\label{Indices}}

Firstly, we consider a portfolio with five asset classes, viz. (i) Hybrid Bond, (ii) Emerging Market, (iii) Commodity, (iv) Bond and (v) Stocks, taken from the R-package `ghyp' (\cite{ghyppack}). The investment timeline consists of  two time periods of study. The first consisting three months of monthly return data (May, June and July of 2008), and the second time period (for August, September and October of 2008). As the second month of period 2 is September 2008, when events like `Bankruptcy of Lehman Brothers' took place, the total volatility during this period is very high. However, for a portfolio manager it is important to identify the different sources of risk contribution from highest to least. 

We consider the same portfolio weights (see table  \ref{PortWeightTable}) during both periods for fair comparison. Here $n=3$ and $p=5$, which implies that regular sample covariance matrix estimator is not stable, as it provides non-unique solution for both periods. We use the Bayesian approach as discussed in section \ref{PostCov}. We select the degrees of freedom for Wishart prior to be $n_0=3.5$, as $(p-n)=~2$, we maintain a balanced prior information by our choice of $c=1.5$  and  choose $\Psi=\lambda^{\prime}\mathrm{I_{p}}$, and $\mathrm{I_{p}}$ is the identity matrix of order $p$, as described in equation (\ref{WishartPrior}), (\ref{WishartPosterior}) and (\ref{EqChooseDf}). We considered simulation size as $N=10000$.

\begin{figure}[h!]
\centering
\caption{Plot showing posterior density of volatility for two periods.\label{post-total-vol}}
\includegraphics[width=6in,height=4in]{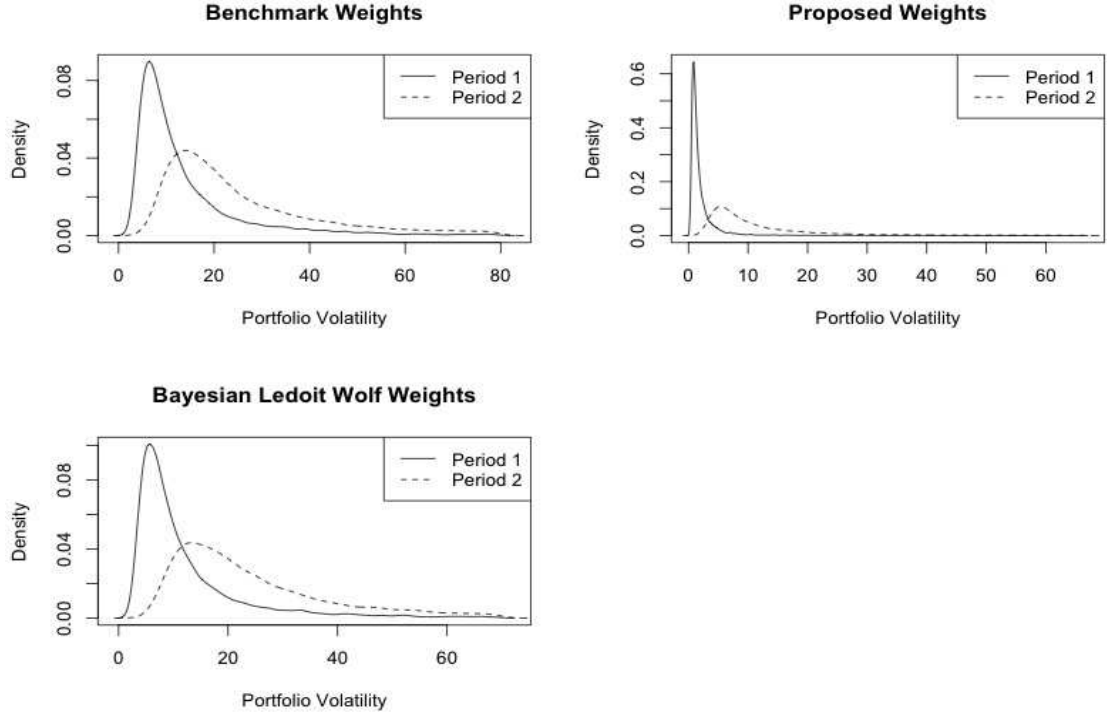}
\end{figure}

We present posterior estimates of total volatility in table \ref{TotalVolTable} and posterior density plot in figure \ref{post-total-vol}. Monthly level total portfolio volatility is estimated by posterior mean and it goes up from $5.18\%$  to $15.21\%$ during first to second period. The portfolio density becomes more positively skewed during the period 2 as presented in figure \ref{post-total-vol}. 

\begin{figure}[h!]
\centering
\caption{Plot showing CCTR for the 2 periods under ad-hoc weights.\label{ad-hoc-cctr}}
\includegraphics[width=6in,height=4in]{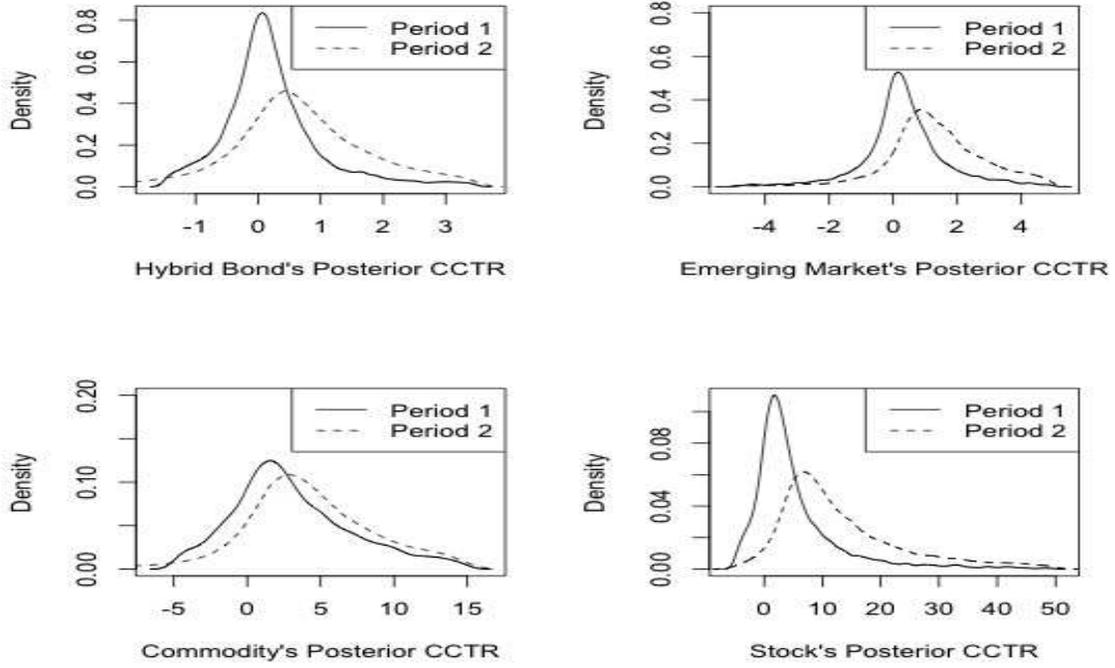}
\end{figure}

\begin{figure}[h!]
\centering
\caption{Plots showing CCTR for the 2 periods under (Bayesian) Ledoit Wolf weights.\label{blw-cctr}}
\includegraphics[width=6in,height=4in]{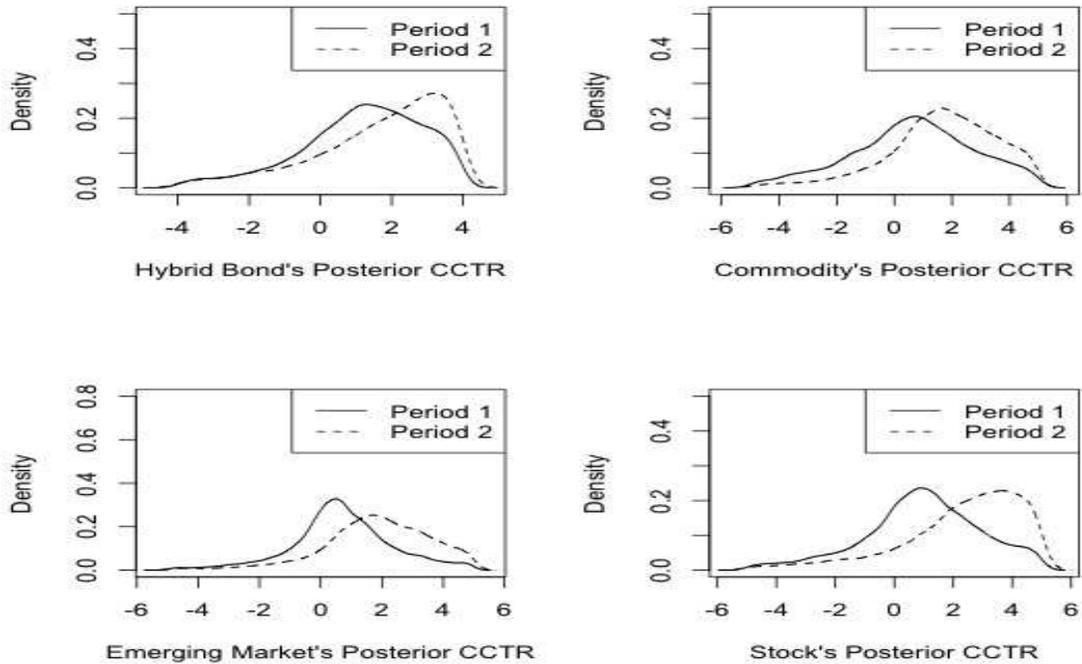}
\end{figure}

\begin{figure}[h!]
\centering
\caption{Plots showing CCTR for the 2 periods under proposed weights.\label{dhd-cctr}}
\includegraphics[width=6in,height=4in]{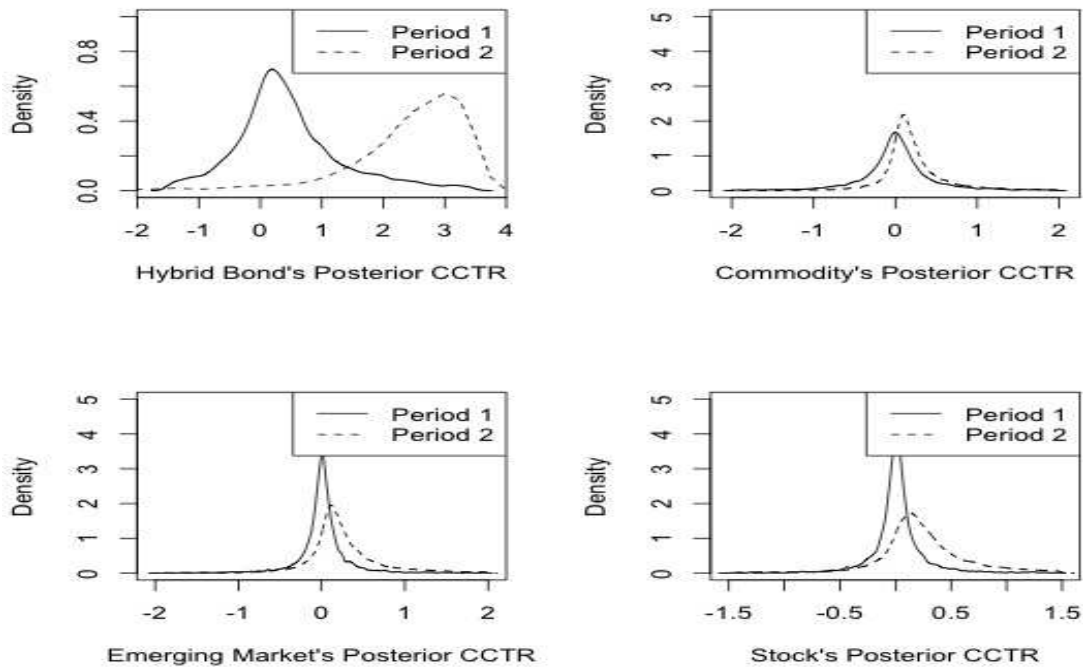}
\end{figure}

\begin{figure}[h!]
\centering
\caption{Plots showing CCTR for the 2 periods for the bond asset under the three weights.\label{bond-cctr}}
\includegraphics[width=6in,height=4in]{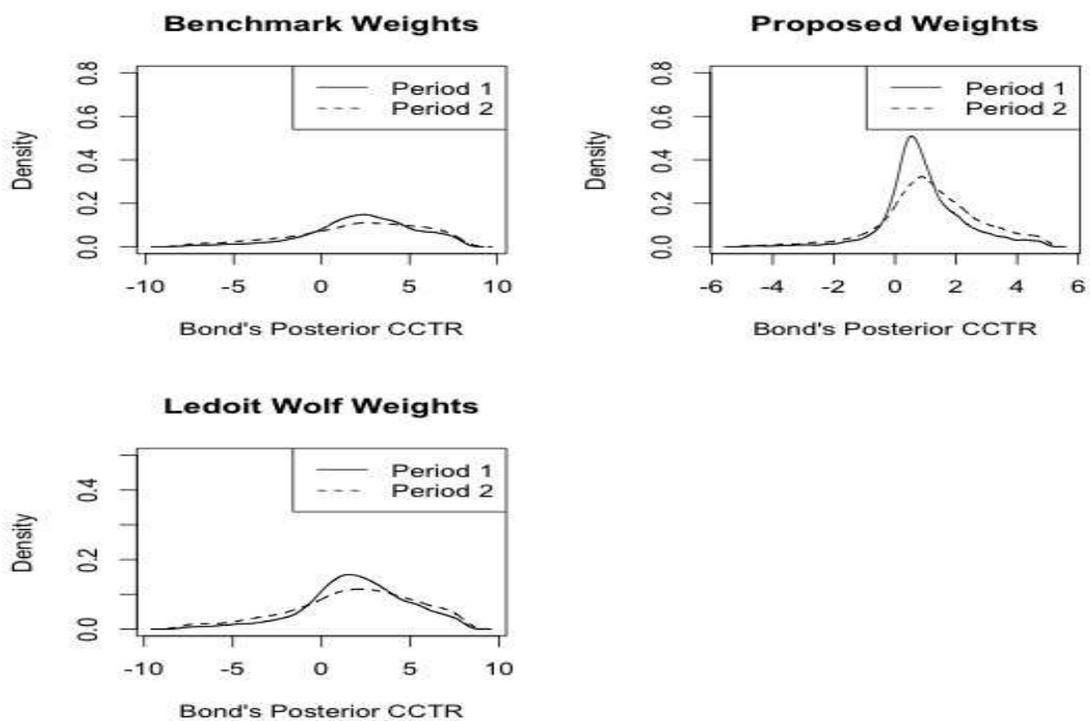}
\end{figure}

We present the posterior density of CCTR ($\zeta$) of five asset classes, viz. (i) Hybrid bond, (ii) Emerging Market, (iii) Commodity, (iv) Bond and (iv) Stock for the three sets of weights in figures \ref{ad-hoc-cctr}, \ref{blw-cctr}, \ref{dhd-cctr} and \ref{bond-cctr}. Clearly we can say from these figures that posterior density of CCTR of four asset classes (viz. Hybrid bond, Emerging Market, Commodity and Stock) have shifted towards right and have become more positively skewed during the second period. It means that these four asset classes have contributed heavily to the total volatility-risk during the second period. Posterior estimates of all five asset classes are presented in tables \ref{PostStatCCTRM}, \ref{PostStatCCTRSD} and \ref{PostStatCCTRCI}. Except `Bond',  for four of the other asset classes,  the posterior mean of CCTR per unit increase in standard deviation  went up significantly during the second period. Therefore we can conclude these four assets classes contributed statistically significantly large amounts towards the total volatility risk of the portfolio. One point to be noted here is that the ad-hoc indicates that among these four asset classes, the posterior mean CCTR of  `Bond' is the least compared to CCTR's of `Emerging Market', `Commodity' and `Stock'. In case of the other two weights namely, BLW (Ledoit Wolf with Bayesian Weights) and DHD (Proposed Weights, in Section (\ref{RiskAna})), the least posterior mean CCTR is presented by `Emerging Market' and `Commodity' respectively. This is explained by noting that the two methods focus mainly on  decreasing volatility (\ref{PortWeightTable}) for the portfolio by optimising the weights accordingly. The allocation of funds is mainly towards `Bond'. This results in a comparitively superior portfolio, with the proposed weight structure (DHD) dominating over the others (\ref{Port-Inf}), showing lower portfolio volatility-risk, maximized return, and consequently a higher Sharpe Ratio in comparision to the others, during the `crisis' period.

However, `Bond' is the only asset class whose posterior density, as presented in figure \ref{bond-cctr},  did not shift towards right. Rather it became more concentrated around zero under the proposed weight structure. Note that `Hybrid Bond' shows the maximum shift to the positive disrection in the second period, contributing greatly to increasing portfolio risk. Also we can see from table \ref{PostStatCCTRCI} that Bayesian $95\%~CI$ for Bond shows a considerable spread to the negative side compared to the other asset classes.

Finally, using the formula as presented in equation (\ref{EqProbCCTR}), we calculated the probability of positive CCTR for both periods and presented the result in table \ref{TabProbCCTR}. Probability of positive CCTR for all the asset classes, except Bond, goes up during period 2. During the same period, the probability of positive CCTR for bond decreases. This showed us that during period 2, which is commonly termed now as a `crisis' period, bond was the only asset class whose volatility-risk showed a decreasing behaviour. This analysis clearly aligns with the intuitive understanding of the bonds market. The advantage of this analysis is that it can provide reasoning to such intuitive understanding in terms of Bayesian probability.

\subsection{National Stock Exchange of India Data} 

Here we consider a market portfolio consisting of equities taken from the NSEI (National Stock Exchange of India) Market. The investment period is fixed at half yearly intervals. Period 1 consists of the (January-June), and Period 2 (July-December), for the years 2005-2014. The data for the respective years is downloaded from the site < https://www.quandl.com/data/NSE?keyword= >. The number of securities in the 10 year time period selected varies from $p=455$ to $p=991$. We consider the daily log-returns for each of the available securities, providing us with $n=125$ days of data, on an average on p securities for both periods. The portfolio weights are obtained using the quadratic-optimisation technique on the data, using regularization procedures listed in Section \ref{PostCov}.

We compare portfolios obtained under the two broad methodologies, that is Ledoit and Wolf (2004a) and the proposed methodolgy as in Section \ref{RiskAna}, using $\Sigma_{23}$, as the regularization matrix. Another comparision is made with the regularization $\Sigma_{13}$ (\ref{lw-weight}), for the sample covariance matrix. Portfolios obtained under the two methods are compared using following basis for comparision (i) Half-Yearly Portfolio Return, (ii) Half-Yearly Portfolio Risk, (iii) Sharpe Ratio and (iv) Portfolio Size. We use the Bayesian weights while conducting the regularization of the covariance matrix. The Wishart prior has the degree of freedom as $n_0=p$, for the respective year. the shrinkage target or prior $\Psi=\lambda^{\prime}\mathrm{I}$, where $\lambda^{\prime}=~ (s_{11},s_{22},\ldots,s_{pp})$, as defined in $\Sigma_{23}$ (\ref{bayes-wt}) in Section \ref{RiskAna}. The simulation size is $N=10000$.

The figures presented in (\ref{return-risk},\ref{esf-var},\ref{sr}) provide a comparative summary for the two methodologies, on the basis of more recent methods of measuring risk, namely `VaR'(Value at Risk) and `ESF'(Expected Shortfall). We present the out-sample performance, including the anticipated return and the volatility of the portfolios over the years. The out-sample performance summary, for the portfolio constructed under the proposed weights is provided in table (\ref{DHD-infer}). In the case of out-sample performances, the period 2007-09 shows a clear dominance of the proposed method, while in other out-sample half-yearly periods it is dominant over the Ledoit and Wolf method, on an average.

We consider the period 2013 (Jul-Dec) in Table (\ref{infer-13-14}) to demonstrate the procedure of inference on a portfolio constructed solely on equity, considering this as the investment period. The same procedure of inference is carried out as in the case of `Indices' dataset, the only difference is in terms of the size of the market portfolio is much larger, making the problem considerably ``more" ill-posed. The shrinkage towards target is more sensitive to $p\to \infty$, while $n$ remains constant, implying $\frac{p}{n} \to \infty$.

Note that from table (\ref{DHD-infer}), the size of the portfolio constructed under the proposed weights utilizes 10$\%$ of the market.We now proceed to examine the stocks common to both portfolios. For a stock with higher CCTR that is, $\mathrm{P}[\zeta>0]$ included in the portfolio constructed using the Ledoit and Wolf weights, presents an even higher $\mathrm{P}[\zeta>0]$ in the portfolio constructed using proposed weights as shown in Table (\ref{infer-13-14}) which provides a major point of difference between the two portfolios, apart from the significant difference in portfolio size. The posterior summary statistics for CCTR of 6 stocks over the two periods are presented in tables (\ref{post-infer-dhd-1h}, \ref{post-infer-dhd-2h}). The stocks X20MICRONS and ABBOTINDIA in the first period (Jul-Dec'13) have a negative CCTR, and contribute to decreasing the volatility-risk of the portfolio in the same period. Comparing this to the second period we see that ABBOTINDIA presents a positive CCTR, showing that the $\mathrm{P}[\zeta>0]$ has increased, which implies increase in portfolio volatility-risk. X20MICRONS preserves a negative CCTR and shows more concentration near the posterior mean. In the second period (Jan-Jun'14) ABGSHIP has a negative CCTR and shows a considerable decrease in CCTR from the first period. The stocks ASTRAL and BAJAJHLDNG show decreased CCTR, their distributions being more concentrated about the posterior mean, in comparision to the first period. The stock BHUSHANSTL, however shows no considerable change over the two periods.

Finally, using the asymptotic weights, instead of the Bayesian weights for the Ledoit Wolf method with the same shrinkage target as indicated in $\Sigma_{13}$ as in (\ref{lw-weight}), we can compute the value for $\rho$, providing us with a comparable, but inferior result in comparision with the proposed portfolio weights. The proposed weights also provide a portfolio of smaller size that lowers transaction cost for the investor.

\section{Conclusion\label{Concl}}

In this paper we discussed about Bayesian approach to regularize the `ill-posed' covariance estimation problem, establishing equivalence of the Bayesian technique with the existing non-parametric techniques. The method also analyzes the sources of risk by estimating the probability of CCTR being positive for any particular security. As CCTR sums up to total volatility, it provides each source's contribution to total volatility. Regular method only estimates CCTR of a source, but it does not estimate the probability of CCTR to be significantly greater (or less) than zero. This paper discussed the methodology to do so. The existing and relatively new measures of risk, like ESF and VaR, are used to analyze the performance of the proposed portfolio weights in comparison to the traditional methods. We presented a parallelizable and easy to implement Monte Carlo method to carry out inference regarding the individual contribution of securities towards total risk. We further presented two empirical studies, the first showed that during the crisis of 2008, a portfolio consisting of five asset classes experienced large volatility risk due to significant increase in the contribution of Stock and Hybrid Bond. During the same period `Bond' was the asset class which contributed least on an average to the risk exposure of the portfolio. Secondly, the performance of the regularization technique under relatively higher dimensions, under computationally efficient Bayesian analysis, to produce effective construction of a portfolio and inference regarding its risk exposure.

\section{Appendix \label{App}}

\textbf{Proof of Result 1:}
\begin{eqnarray*}
\mathrm{tr}(\Sigma^2)-\frac{[\mathrm{tr}(\Sigma)]^2}{p} &=& \sum_{i=1}^{p} \sum_{j=1}^{p} \sigma_{ij}^2 - \frac{1}{p}\Big[ \sum_{i=1}^{p} \sigma_{ii}\Big]^2\\
&=& \frac{1}{p}\Big[ p\sum_{i=1}^{p} \sigma_{ii}^2 +p\mathop{\sum \sum}_{i\neq j}^{p} \sigma_{ij}^2- \Big\lbrace \sum_{i=1}^{p} \sigma_{ii}^2 +\mathop{\sum \sum}_{i\neq j}^{p} \sigma_{ii}\sigma_{jj}\Big\rbrace\Big]\\
&=& \frac{1}{p}\Big[ (p-1)\sum_{i=1}^{p} \sigma_{ii}^2+\mathop{\sum \sum}_{i\neq j}^{p} (p\; \sigma_{ij}^2-\sigma_{ii}\sigma_{jj})\Big]\\
&=& \frac{1}{p}\Big[(p-1)\sum_{i=1}^{p} \sigma_{ii}^2+\mathop{\sum \sum}_{i\neq j}^{p} \sigma_{ii} \sigma_{jj}(p\; \rho_{ij}^2-1)\Big].
\end{eqnarray*}
Consequently we have, 
$$
\alpha^2=\frac{1}{p^2}\Big[(p-1)\sum_{i=1}^{p}\sigma_{ii}^2+\mathop{\sum \sum}_{i\neq j}^{p}\sigma_{ii}\sigma_{jj}(p\;\rho_{ij}^2-1)\Big].
$$
\textbf{Proof of Result 2:}\newline
Note that, the following result is true for the covariance matrix, S,
\begin{eqnarray*}
\mathrm{E}[\mathrm{tr}(S^2)]&=&\mathrm{tr}[\mathrm{E}(S^2)]\\
&=& \mathrm{tr}\Big[\mathrm{Var}(S)+[\mathrm{E}(S)]^2 \Big]\\
&=& \mathrm{tr}\Big[(n-1)((\sigma_{ij}^2+\sigma_{ii}\sigma_{jj}))_{i=1\ldots p,j=1\ldots p}+(n-1)^2\Sigma^2 \Big]\\
&=& (n-1)\Big[2\sum_{i=1}^{p}\sigma_{ii}^2+(n-1)\; \mathrm{tr}(\Sigma^2)\Big].
\end{eqnarray*}
Using the result obtained above we have,
\begin{eqnarray*}
\beta^2&=&\mathrm{E}\Big[\Big| \Big| \frac{S}{n-1}-\Sigma\Big| \Big|^2 \Big] \\
&=& \frac{1}{p} \mathrm{E}\Big[\mathrm{tr}\Big\lbrace \Big(\frac{S}{n-1} -\Sigma \Big)\Big(\frac{S}{n-1} -\Sigma \Big)^{\prime}\Big\rbrace \Big]\\
&=&\frac{1}{p} \mathrm{E}\Big[\mathrm{tr} \Big(\frac{S^2}{(n-1)^2}\Big) -2 \mathrm{tr}\Big(\frac{S}{n-1}\Sigma \Big) +\mathrm{tr} \Big(\Sigma^{2}\Big)\Big\rbrace \Big].\\
\end{eqnarray*}
Now using the fact that $\mathrm{E}[\mathrm{tr}(.)]=\; \mathrm{tr}[\mathrm{E}(.)]$, $S \sim \mathcal{W}(n-1,\Sigma) \Rightarrow \mathrm{E}[S]=\; (n-1)\Sigma$, and the result above, we have
\begin{eqnarray*}
\beta^2&=& \frac{1}{p} \mathrm{tr}\Big[\mathrm{E} \Big(\frac{S^2}{(n-1)^2}\Big) -2 \mathrm{E}\Big(\frac{S}{n-1}\Sigma \Big) +\mathrm{E} \Big(\Sigma^{2}\Big)\Big\rbrace \Big]\\
&=& \frac{1}{p} \Big[ \frac{(n-1)\lbrace 2\sum_{i=1}^{p} \sigma_{ii}^2 + (n-1) \mathrm{tr}(\Sigma^2)\rbrace}{(n-1)^2}-\frac{2}{n-1}(n-1)\mathrm{tr}(\Sigma^2)+\mathrm{tr}(\Sigma^2)\Big]\\
&=&\frac{2}{n-1}\Big[ \frac{1}{p}\sum_{i=1}^{p}\sigma_{ii}^2\Big].
\end{eqnarray*}
\textbf{Proof of Result 3: }
\begin{eqnarray*}
\delta^2&=& \mathrm{E}\Big[\Big| \Big| \frac{S}{n-1}-\mu\mathrm{I} \Big| \Big|^2 \Big]\\
&=& \frac{1}{p}\mathrm{E}\Big[ \mathrm{tr} \Big\lbrace \Big(\frac{S}{n-1}-\mu\mathrm{I}\Big)\Big(\frac{S}{n-1}-\mu\mathrm{I}\Big)^{\prime}\Big\rbrace\Big]\\
&=& \frac{1}{p}\mathrm{E}\Big[\mathrm{tr} \Big(\frac{S^2}{(n-1)^2}\Big)-\frac{2\mu}{n-1}\mathrm{tr}(S)+\mu^2\mathrm{tr}\mathrm{I}\Big]\\
&=& \frac{1}{p}\mathrm{tr}\Big[ \mathrm{E}\Big(\frac{S^2}{(n-1)^2}\Big)-\frac{2\mu}{n-1}(n-1)\Sigma+p\mu^2\Big]\\
&=& \frac{1}{p}\Big[ \frac{(n-1)\lbrace 2\sum_{i=1}^{p} \sigma_{ii}^2 + (n-1) \mathrm{tr}(\Sigma^2)\rbrace}{(n-1)^2}-2\frac{[\mathrm{tr}(\Sigma)]^2}{p}+\frac{[\mathrm{tr}(\Sigma)]^2}{p} \Big]\\
&=& \frac{2}{n-1}\Big[ \frac{1}{p}\sum_{i=1}^{p}\sigma_{ii}^2\Big] + \frac{1}{p}\Big\lbrace\mathrm{tr}(\Sigma^2) -\frac{[\mathrm{tr}(\Sigma)]^2}{p}\Big\rbrace=\; \alpha^2+\beta^2.
\end{eqnarray*}
\textbf{Proof of Result 5: }
\begin{eqnarray*}
\rho &=& \mathop{\mathrm{argmin}}_{\rho \geq 0} \mathrm{E}[L(\rho)] ~=~ \mathop{\mathrm{argmin}}_{\rho \geq 0} \Big|\Big|\Sigma_3-\Sigma\Big|\Big|^2\\
&=&  \mathop{\mathrm{argmin}}_{\rho \geq 0} \Big|\Big|\rho \lambda^{\prime}\mathrm{I_{p}}+(1-\rho)\frac{S}{n-1}-\Sigma\Big|\Big|^2~=~\mathop{\mathrm{argmin}}_{\rho \geq 0} \Big|\Big|\rho \mathop{\mathrm{Diag}}_{i}~\lbrace s_{11},s_{22},\ldots,s_{pp}\rbrace+(1-\rho)\frac{S}{n-1}-\Sigma\Big|\Big|^2.\\
\end{eqnarray*}
Now considering $L(\rho)~=~\Big|\Big|\rho \mathop{\mathrm{Diag}}_{i}~\lbrace s_{11},s_{22},\ldots,s_{pp}\rbrace+(1-\rho)\frac{S}{n-1}-\Sigma\Big|\Big|^2$ we have,
\begin{eqnarray*}
\mathrm{E}[L(\rho)]&=& \frac{1}{p}\mathrm{E}\Big[\mathrm{tr}\Big\lbrace\Big(\rho \mathop{\mathrm{Diag}}_{i}~\lbrace s_{11},s_{22},\ldots,s_{pp}\rbrace+(1-\rho)\frac{S}{n-1}-\Sigma\Big)\Big(\rho \mathop{\mathrm{Diag}}_{i}~\lbrace s_{11},s_{22},\ldots,s_{pp}\rbrace+(1-\rho)\frac{S}{n-1}-\Sigma\Big)^{\prime}\Big\rbrace\Big]\\
&=& \frac{1}{p} \mathrm{tr} \Big(\mathrm{E}\Big[\rho^2\mathop{\mathrm{Diag}}_{i}~\lbrace s_{ii}\rbrace^2+2\rho(1-\rho)\mathop{\mathrm{Diag}}_{i}~\lbrace s_{ii}\rbrace\frac{S}{n-1}-2\rho\mathop{\mathrm{Diag}}_{i}~\lbrace s_{ii}\rbrace\Sigma-2(1-\rho)\Sigma\frac{S}{n-1}+\Sigma^2\Big]\Big).\\
\end{eqnarray*}
Since, $S\sim\mathcal{W}(n-1,\Sigma)$ implies, $\mathrm{E}[S]=(n-1)\Sigma$, we have
\begin{eqnarray*}
\mathrm{E}[L(\rho)] &=& \frac{1}{p}\Big[\sum_{i=1}^{p}(n-1)\rho^2(2\sigma_{ii}^2+\sigma_{ii})+2\sum_{i=1}^{p}\rho(1-\rho)(2\sigma_{ii}^2+\sigma_{ii})-2\sum_{i=1}^{p}\rho(n-1)\sigma_{ii}^2+\sum_{i=1}^{p}\frac{(1-\rho)^2}{n-1}(2\sigma_{ii}^2+\sigma_{ii})\Big].
\end{eqnarray*}
Re-arranging the terms in the above equation we have
\begin{eqnarray*}
\mathrm{E}[L(\rho)] &=& \frac{1}{p}\Big[2\Big\lbrace\Big(\sqrt{n-1}\rho+\frac{1-\rho}{\sqrt{n-1}}\Big)^2-\rho(n-1)\Big\rbrace\sum_{i=1}^{p}\sigma_{ii}^2+\Big(\sqrt{n-1}\rho+\frac{1-\rho}{\sqrt{n-1}}\Big)^2\sum_{i=1}^{p}\sigma_{ii}+(2\rho-1)\sum_{i=1}^{p}\sum_{j=1}^{p}\sigma_{ij}^2\Big].
\end{eqnarray*}
If we denote $R(\rho)=\mathrm{E}[L(\rho)]$, then to minimize $R(\rho)$ we have the normal equation, by differentating w.r.t $\rho$, and rearranging terms, for co-efficient of $\rho$
\begin{eqnarray*}
\frac{\partial R(\rho)}{\partial\rho}&=&0\\
\Rightarrow ~ (n-2)^2[2\sum_{i=1}^{p}\sigma_{ii}^2+\sum_{i=1}^{p}\sigma_{ii}]\rho &=& [(n-2)^2+1]\sum_{i=1}^{p}\sigma_{ii}^2-(n-2)\sum_{i=1}^{p}\sum_{i=1}^{p}\sigma_{ii}-(n-1)\sum_{i=1}^{p}\sum_{j=1}^{p}\sigma_{ij}^2,\\
\end{eqnarray*}
which is the required result.

\newpage

\section{Tables and Figures}

\begin{table}[ht]
\centering
\caption{Portfolio weights for five asset classes  \label{PortWeightTable}}

\begin{tabular}{|r|r|r|r|r|r|}
  \hline
  & Hybrid Bond & Emerging Mkt & Commodity & Bond & Stock \\ 
  \hline
  ad-hoc Weight & $5\%$ & $5\%$ & $20\%$  & $40\%$ & $30\%$ \\ 
  Ledoit Wolf & $28\%$ & $9\%$ & $13\%$ & $33\%$ & $18\%$ \\ 
  Markowitz & $28\%$ & $0.7\%$ & $1\%$ & $70\%$ & $0.8\%$ \\ 
   \hline
\end{tabular}
\end{table}

\begin{table}[h!]
\centering
\caption{Table showing the Prob($\zeta>0$)  for all five asset class over two periods\label{TabProbCCTR}}

\begin{tabular}{|r||r|r||r|r||r|r|}
  \hline
  Asset & Period 1 & Period 2 & Period 1 & Period 2 & Period 1 & Period 2 \\ 
  \hline
  Method & BM & BM &BLW &BLW &DHD &DHD \\\hline\hline
  Hy.bond & 0.57 & 0.77 & 0.80 & 0.90 & 0.70 & 0.99 \\ 
  Emerging.mkt & 0.64 & 0.87 & 0.71 & 0.89 & 0.53 & 0.81 \\ 
  Commodity & 0.74 & 0.87 & 0.66 & 0.86 & 0.51 & 0.80 \\ 
  Bond & 0.83 & 0.77 & 0.80 & 0.75 & 0.84 & 0.83 \\ 
  Stock & 0.77 & 0.94 & 0.73 & 0.92 & 0.54 & 0.82 \\
   \hline
\end{tabular}
\end{table}

\begin{table}[h!]
\centering
\caption{Posterior Statistics of Total Portfolio Volatility \label{TotalVolTable}}

\begin{tabular}{|c|c|c|c|}
\hline
Asset Class  & Posterior Statistics & Period 1          & Period 2      \\ \hline
Portfolio    &  Mean (SD)           & 18.77 (56.18)     & 43.12 (156.84)  \\
(ad-hoc)  &  $95\% ~~CI$         & (3.78 , 80.30)    & (8.62 , 183.76)  \\\hline
Portfolio    &  Mean (SD)           & 16.71 (46.24)     & 39.76 (113.01)  \\
(BLW)        &  $95\% ~~CI$         & (3.35 , 71.72)    & (8.02 , 166.74)  \\\hline
Portfolio    &  Mean (SD)           & 2.36 (5.57)       & 15.86 (49.95)  \\
(DHD)        &  $95\% ~~CI$         & (0.48 , 10.40)    & (3.16 , 66.35)  \\\hline
\end{tabular}
\end{table}

\begin{table}[ht]
\centering
\caption{Posterior Statistics of CCTR five asset classes: Mean\label{PostStatCCTRM}}

\begin{tabular}{|r|r|r|r|r|r||r|r|r|r|r|r|}
  \hline
  Method(P) & HB & EM. & CM & Bo & St & Method(P) & HB & EM. & CM & Bo & St \\ 
  \hline
  BM(I) & 0.22 & 0.62 & 4.76 & 7.33 & 5.84 &BM(II)& 1.49 & 3.33 & 9.56 & 1.02 & 26.22 \\ 
  BLW(I) & 4.57 & 1.46 & 2.01 & 5.52 & 3.15 &BLW(II)& 10.96 & 4.69 & 5.65 & 7.21 & 11.26 \\ 
  DHD(I) & 0.75 & 0.03 & 0.01 & 1.56 & 0.03 &DHD(II)& 11.97 & 0.41 & 0.33 & 2.69 & 0.45 \\ 
   \hline
\end{tabular}
\end{table}

\begin{table}[h!]
\centering
\caption{Posterior Statistics of CCTR five asset classes: SD\label{PostStatCCTRSD}}

\begin{tabular}{|r|r|r|r|r|r||r|r|r|r|r|r|}
  \hline
  Method(P) & HB & EM. & CM & Bo & St &Method(P)& HB & EM. & CM & Bo & St  \\ 
  \hline
  BM(I) & 4.87 & 5.91 & 33.79 & 36.78 & 34.42 &BM(II)& 7.14 & 19.79 & 48.46 & 6.78 & 77.43 \\ 
  BLW(I) & 26.56 & 10.48 & 22.72 & 30.11 & 20.17 &BLW(II)& 39.20 & 39.59 & 36.43 & 46.47 & 72.66 \\ 
  DHD(I) & 4.98 & 0.87 & 6.61 & 5.45 & 0.72 &DHD(II)& 39.14 & 2.75 & 2.34 & 11.72 & 2.09 \\ 
   \hline
\end{tabular}
\end{table}

\begin{table}[h!]
\centering
\caption{Posterior Statistics of CCTR five asset classes: $95\%$ CI\label{PostStatCCTRCI}}

\begin{tabular}{|r|r|r|r|r|r||r|r|r|r|r|r|}
  \hline
  Method(P) & HB & EM. & CM & Bo & St & Method(P) & HB & EM. & CM & Bo & St\\  
  \hline
  BM(I) 2.5\% & -3.20 & -4.37 & -16.84 & -11.41 & -15.15 &BM(II) 2.5\%& -3.51 & -3.13 & -9.11 & -23.89 & -6.02 \\ 
  BM(I) 97.5\% & 4.16 & 7.01 & 40.30 & 44.77 & 41.64 &BM(II) 97.5\%& 9.83 & 15.70 & 48.63 & 68.64 & 94.49 \\ 
  BLW(I) 2.5\% & -8.82 & -6.61 & -14.24 & -10.84 & -10.08 &BLW(II) 2.5\%& -9.07 & -4.81 & -6.93 & -21.15 & -5.82 \\ 
  BLW(I) 97.5\% & 29.50 & 13.45 & 24.72 & 36.48 & 24.16 &BLW(II) 97.5\%& 61.70 & 28.60 & 32.37 & 54.49 & 53.44 \\ 
  DHD(I) 2.5\% & -2.84 & -0.82 & -1.73 & -2.32 & -0.66 &DHD(II) 2.5\%& 1.26 & -0.65 & -0.63 & -4.76 & -0.74 \\ 
  DHD(I) 97.5\% & 6.04 & 0.85 & 1.76 & 10.33 & 0.78 &DHD(II) 97.5\%& 51.47 & 2.50 & 2.13 & 17.94 & 2.91  \\ 
   \hline
\end{tabular}
\end{table}

\begin{table}[h!]
\centering
\caption{Portfolio Inference under the three methods\label{Port-Inf}}
\begin{tabular}{|r|r|r|r|}
  \hline
 Method & Proposed Weights & BLW & ad-hoc\\ 
  \hline
  Volatility & 0.03 & 0.07 & 0.07 \\ 
  Mean & -0.07 & -0.23 & -0.24 \\ 
  Sharpe Ratio & -2.59 & -3.31 & -3.26 \\ 
   \hline
\end{tabular}
\end{table}

%\begin{figure}[h!]
%\centering
%\caption{Plot showing out-sample anticipated return and Risk.\label{return-risk}}
%\subfloat[return out-sample]{\includegraphics[width=3.5in,height=3in]{Fig-ret-outsample}}
%\subfloat[Risk out-sample]{\includegraphics[width=3.5in,height=3in]{Fig-risk-outsample}}
%\end{figure}

%\begin{figure}[h!]
%\centering
%\caption{Plot showing out-sample ESF and VaR.\label{esf-var}}
%\subfloat[ESF out-sample]{\includegraphics[width=3.5in,height=3in]{Fig-esf-outsample}}
%\subfloat[VaR out-sample]{\includegraphics[width=3.5in,height=3in]{Fig-var-outsample}}
%\end{figure}

%\begin{figure}[h!]
%\centering
%\caption{Plot showing out-sample Sharpe Ratio.\label{sr}}
%\includegraphics[width=3.5in,height=3in]{Fig-SR-outsample}
%\end{figure}

%\newpage 

\begin{table}[h!]
\centering
\caption{Table showing Posterior Inference on CCTR NSEI Data for Jul-Dec'13.\label{post-infer-dhd-1h}}

\begin{tabular}{|r|r|r|r|r|r|r|}
  \hline
  Equity Listing & X20MICRONS & ABBOTINDIA & ABGSHIP & ASTRAL & BAJAJHLDNG & BHUSANSTL \\ 
  \hline
  Posterior Mean & -0.14 & -0.01 & 0.10 & 1.12 & 1.20 & 0.36 \\ 
  Posterior SD & 4.88 & 2.16 & 4.46 & 35.89 & 37.61 & 16.53 \\ 
  Posterior 2.5\% & -3.94 & -1.65 & -4.14 & -29.02 & -23.86 & -13.96 \\ 
  Posterior 97.5\% & 3.68 & 1.81 & 4.55 & 34.58 & 30.75 & 15.86 \\ 
   \hline
\end{tabular}
\end{table}

\begin{table}[h!]
\centering
\caption{Table showing Posterior Inference on CCTR NSEI Data for Jan-Jun '14.\label{post-infer-dhd-2h}}

\begin{tabular}{|r|r|r|r|r|r|r|}
  \hline
  Equity Listing & X20MICRONS & ABBOTINDIA & ABGSHIP & ASTRAL & BAJAJHLDNG & BHUSANSTL \\ 
  \hline
  Posterior Mean & -0.02 & 0.02 & -0.05 & 0.05 & 0.08 & 0.39 \\ 
  Posterior SD & 0.88 & 4.70 & 4.55 & 2.88 & 2.90 & 18.35 \\ 
  Posterior 2.5\% & -0.62 & -3.87 & -4.03 & -1.42 & -1.87 & -6.54 \\ 
  Posterior 97.5\% & 0.58 & 3.73 & 3.84 & 1.51 & 2.20 & 7.79 \\ 
   \hline
\end{tabular}
\end{table}

\begin{table}[ht]
\centering
\caption{Table showing out-sample performance of Proposed Weights\label{DHD-infer}}

\begin{tabular}{|r|r|r|r|r|r|}
  \hline
  Period & Portfolio Return & Portfolio Risk & Sharpe Ratio & Portfolio Size & Market Size \\ 
  \hline
  2005 (Jul-Dec) & 10.12 & 16.29 & 0.62 & 55.00 & 455.00 \\ 
  2006 (Jan-Jun) & -29.09 & 23.14 & -1.26 & 64.00 & 513.00 \\ 
  2006 (Jul-Dec) & 80.25 & 8.42 & 9.53 & 58.00 & 513.00 \\ 
  2007 (Jan-Jun) & 1.34 & 8.76 & 0.15 & 99.00 & 659.00 \\ 
  2007 (Jul-Dec) & 66.97 & 11.80 & 5.68 & 82.00 & 659.00 \\ 
  2008 (Jan-Jun) & -51.49 & 20.66 & -2.49 & 77.00 & 702.00 \\ 
  2008 (Jul-Dec) & -30.13 & 16.78 & -1.80 & 48.00 & 702.00 \\ 
  2009 (Jan-Jun) & 3.59 & 5.33 & 0.67 & 39.00 & 293.00 \\ 
  2009 (Jul-Dec) & 17.79 & 16.21 & 1.10 & 38.00 & 293.00 \\ 
  2010 (Jan-Jun) & -8.95 & 14.65 & -0.61 & 64.00 & 944.00 \\ 
  2010 (Jul-Dec) & -13.93 & 12.32 & -1.13 & 84.00 & 944.00 \\ 
  2011 (Jan-Jun) & -13.85 & 11.02 & -1.26 & 98.00 & 966.00 \\ 
  2011 (Jul-Dec) & -39.55 & 10.88 & -3.64 & 75.00 & 966.00 \\ 
  2012 (Jan-Jun) & -2.10 & 5.34 & -0.39 & 76.00 & 895.00 \\ 
  2012 (Jul-Dec) & 6.77 & 5.00 & 1.35 & 110.00 & 895.00 \\ 
  2013 (Jan-Jun) & -11.22 & 5.85 & -1.92 & 148.00 & 930.00 \\ 
  2013 (Jul-Dec) & -0.61 & 4.07 & -0.15 & 87.00 & 930.00 \\ 
  2014 (Jan-Jun) & 0.37 & 11.08 & 0.03 & 94.00 & 991.00 \\ 
   \hline
\end{tabular}
\end{table}

\begin{table}[h!]
\centering
\caption{Table showing the Prob($\zeta>0$) for the common stocks in the two methodologies \label{infer-13-14}}
\begin{tabular}{|r|r|r|}
  \hline
  Equity & Period 1 (Jul-Dec'13) & Period 1 (Jul-Dec'13) \\ 
  \hline
  Method & BLW & DHD \\\hline
  DBSTOCKBRO & 0.55  & 0.83  \\ 
  SWANENERGY & 0.55  & 0.57  \\ 
  SEINV & 0.54  & 0.59  \\ 
  SHARONBIO & 0.57 & 0.74  \\ 
  PFIZER & 0.56 & 0.60  \\ 
  NESCO & 0.56 & 0.55  \\ 
   \hline
\end{tabular}
\end{table}


\begin{thebibliography}{10}

\bibitem{Anderson1984}
T.~W. Anderson.
\newblock {\em An Introduction to Multivariate Statistical Analysis 2 Ed.}
\newblock Wiley, 1984.

\bibitem{Baigent2014}
G.~G. Baigent.
\newblock X-sigma-rho and market efficiency.
\newblock {\em International Journal of Finance and Banking}, 1(1):39--43,
  2014.

\bibitem{Bapat1989}
R.~B. Bapat.
\newblock Infinite dividibility of multivariate gamma distributions and
  m-matrices.
\newblock {\em Sankhya: The Indian Journal of Statistics}, 51(1):73--78, 1989.

\bibitem{DasDey2010}
S.~Das and D.~K. Dey.
\newblock On bayesian inference for generalized multivariate gamma
  distribution.
\newblock {\em Statistics and Probability Letters}, 80:1492--1499, 2010.

\bibitem{Knuth1976}
Knuth E.~Graham.
\newblock Big o micron and big omega and big theta.
\newblock {\em SIGACT News}, pages 18--24, 1976.

\bibitem{Gelman2004}
Andrew Gelman, John~B. Carlin, Hal~S. Stern, David~B. Dunson, Aki Vehtari, and
  Donald~B. Rubin.
\newblock {\em Bayesian Data Analysis}.
\newblock Chapman and Hall/CRC, 2004.

\bibitem{GolosnoyOkhrin2007}
V.~Golosnoy and Y.~Okhrin.
\newblock Multivariate shrinkage for optimal portfolio weights.
\newblock {\em The European Journal of Finance}, 13:441--458, 2007.

\bibitem{LedoitWolf2003}
O.~Ledoit and M~Wolf.
\newblock Improved estimation of the covariance matrix of stock returns with an
  application to portfolio selection.
\newblock {\em Journal of Empirical Finance}, 10(5):603--621, 2003.

\bibitem{LedoitWolf2004a}
O.~Ledoit and M~Wolf.
\newblock A well-conditioned estimator for large-dimensional covariance
  matrices.
\newblock {\em Journal of Multivariate Analysis}, 88:365--411, 2004a.

\bibitem{LedoitWolf2004b}
O.~Ledoit and M~Wolf.
\newblock Honey, i shrunk the sample covariance matrix.
\newblock {\em Journal of Portfolio Management}, 30(4):110--119, 2004b.

\bibitem{ghyppack}
David Luethi and Wolfgang Breymann.
\newblock ghyp: A package on the generalized hyperbolic distribution and its
  special cases, 2013.
\newblock R package v 1.5.6.

\bibitem{Markowitz1952}
Harry~M. Markowitz.
\newblock Portfolio selection.
\newblock {\em Journal of Finance}, pages 77--91, 1952.

\bibitem{Matloff2011}
N.~Matloff.
\newblock The art of r programming: A tour of statistical software design.
\newblock {\em The Journal of Portfolio Management}, page 347, 2011.

\bibitem{MencherDavis2011}
J.~Menchero and B.~Davis.
\newblock Risk contribution is exposure time's volatility time's correlation:
  Decomposing risk using the x-sigma-rho formula.
\newblock {\em The Journal of Portfolio Management}, 37(2):97--106, 1989.

\bibitem{Fabozzi2008}
Svetlozar~T. Rachev, John S.~J. Hsu, and Biliana S.
\newblock {\em Bayesian Methods in Finance (Frank J. Fabozzi Series)}.
\newblock Wiley, 2008.

\bibitem{DavidRuppert2004}
D.~Ruppert.
\newblock {\em Statistics and Finance An Introduction}.
\newblock Springer, 2004.

\end{thebibliography}
\end{document}